\documentclass[a4papers]{article}
\usepackage{amssymb,amsthm,amsmath,authblk}
\numberwithin{equation}{section}

\begin{document}

\title{Cumulative Distance Enumerators of Random Codes
and their Thresholds}
\author{\small Yun Fan$^{\bf 1}$,~ San Ling$^{\bf 2}$,~
        Hongwei Liu$^{\bf 1}$,~ Jing Shen$^{\bf 3}$,~
        Chaoping Xing$^{\bf 2}$}
\affil{\small
 $^{\bf 1}$ School of Mathematics and Statistics,
 Central China Normal University, Wuhan 430079, China\\
  $^{\bf 2}$ School of Physical \& Mathematical Sciences,
  Nanyang Technological University, Singapore 637616, Singapore\\
 $^{\bf 3}$ College of Science, Naval University of Engineering,
  Wuhan, 430033, China
}
\date{}%\date{Sept 2011}
\maketitle

\insert\footins{\footnotesize{\it Email addresses}:
 yunfan02@yahoo.com.cn (Yun Fan); lingsan@ntu.edu.sg (San Ling);
 h\_w\_liu@yahoo.com.cn (Hongwei Liu);
 shendina@hotmail.com (Jing Shen);
 xingcp@ntu.edu.sg (Chaoping Xing).
}

\begin{abstract}
Cumulative weight enumerators of random linear codes
are introduced, their asymptotic properties are studied,
and very sharp thresholds are exhibited;
as a consequence, it is shown that
the asymptotic Gilbert-Varshamov bound is a very sharp
threshold point for the density of the linear codes
whose relative distance is greater than a given positive number.
For arbitrary random codes, similar settings and results are
exhibited; in particular, the very sharp threshold point for
the density of the codes whose relative distance
is greater than a given positive number is located
at half the asymptotic Gilbert-Varshamov bound.
\end{abstract}

\section{Introduction}

Let $F$ be an alphabet for coding with cardinality $|F|=q>1$.
Any non-empty subset $C$ of $F^n$ is called a code of length $n$;
the fraction $R(C)=\frac{\log_q|C|}{n}$ is said to be
the {\em rate} of the code~$C$. Codewords,
i.e. the elements of $C$, may be changed at some probability
due to noise when they are communicated through a channel.
Shannon \cite{S} introduced the {\em capacity} of the channel
and showed that: if a number $r$ is below the capacity then
there are codes of rate $r$ and decoding rules
such that the probability of recovering the codewords goes
to~$1$ as $n\to\infty$; however, if $r$ is beyond the capacity
then for any code of rate $r$ the probability of recovering the
codewords is bounded from above by a number less than~$1$.
For the latter case, Wolfowitz \cite{W} further showed that
for any code of rate $r$ the probability of recovering the
codewords goes to $0$ as $n\to\infty$.
Following terminology from physics, Shannon's capacity is
a {\em phase transition point}, also known as a {\em threshold point},
for the channel from high fidelity to no fidelity.

Hamming \cite{H} introduced the
{\em Hamming distance} to construct error-correcting codes.
We denote the {\em minimum Hamming distance} of any code
$C\subseteq F^n$ by $d_H(C)$;
denote the fraction $\frac{d_H(C)}{n}$ by $\Delta(C)$,
and call it the {\em relative distance} of~$C$.
%The larger both the relative distance and the rate,
%the more powerful the code is.
An important question is: how to get the highest rate
of the codes in~$F^n$ with relative distance
greater than a given positive number~$\delta$?
Gilbert~\cite{Gil} and Varshamov \cite{V} exhibited
a function in $\delta$ and $n$,
called the {\em Gilbert-Varshamov bound},
to bound from below this highest rate, cf. \cite[\S2.8, \S2.9]{HP}.
The limit as ${n\to\infty}$ of this lower bound,
which we denote by ${\rm GV}_q(\delta)$,
is called the {\em asymptotic Gilbert-Varshamov bound},
or {\em GV-bound} in short.
The GV-bound coincides exactly with Shannon's
capacity of the {\em $q$-ary symmetric channel},
so it is also called the {\em Gilbert-Shannon-Varshamov bound}.
When $F$ is a finite field, there have been several families of
linear codes in \cite{TVZ,JV,VW,GZ} whose parameters improve
the Gilbert-Varshamov bound (as a function in $\delta$ and $n$).
In the asymptotic sense, the family in \cite{TVZ} does
exceed the GV-bound but only for the case when $q\ge 49$.

For a random linear code $L_r$ of rate $r$,
Varshamov in \cite{V} proved that: if $r<{\rm GV}_q(\delta)$,
then the probability of the relative distance of $L_r$
being greater than $\delta$,
denoted by $\Pr\big(\Delta(L_r)>\delta\big)$,
goes to~$1$ as $n\to\infty$.
Note that the relative distance $\Delta(L_r)$ of the
random linear code $L_r$ of rate $r$ is a random variable.
Pierce \cite{P} proved that $\Delta(L_r)$ is asymptotically
distributed at one point $\delta_{\rm GV}(r)$,
where $\delta_{\rm GV}(r)$ denotes the solution of the equation
$r={\rm GV}_q(x)$ in the variable $x$ and is called
the {\em GV-distance}. From this result of Pierce,
it could further follow that: if $r>{\rm GV}_q(\delta)$ then
$\Pr\big(\Delta(L_r)>\delta\big)$ goes to $0$ as $n\to\infty$.
In other words, the GV-bound ${\rm GV}_q(\delta)$
is a threshold point from high density to extreme sparsity
of the linear codes of rate $r$ whose relative distance
is greater than a given $\delta$.

For a binary random code $C$ from a
so-called {\em random code ensemble},
the {\em distance enumerator with respect to a given codeword},
i.e. the number of the codewords with a given distance
from the given codeword, has been extensively studied;
cf. \cite{Gal}, \cite{Mon}, \cite{M}.
Barg and Forney~\cite{BF} investigated,
by means of Chebychev's inequality,
the {\em distance enumerator} and showed that
the relative distance $\Delta(C)$ is asymptotically distributed
at the point $\delta_{\rm GV}(2r)$.
Similar to the linear case but not as precisely as what
Pierce \cite{P} did, it could follow that
$\frac{1}{2}{\rm GV}_q(\delta)$ is a threshold point
for the density of the codes of rate $r$ whose relative
distance is greater than a given $\delta$.

In this paper we consider a more general setting
with a more sensitive second moment method, including the
Cauchy inequality and the Paley-Zygmund inequality (see \cite{PZ}).
For a random linear code~$L$ of length $n$ and rate $r$ we
define the {\em cumulative weight enumerator} to be the
number of codewords whose weight is bounded from above,
and study its asymptotic property.
We exhibit very sharp threshold points
for the distributions of the cumulative weight enumerator.
It is just a special case of the thresholds we obtain that
the GV-bound ${\rm GV}_q(\delta)$ is a very sharp threshold point
for the density of the linear codes of rate $r$
whose relative distance is greater than a given $\delta$.
Furthermore, the result of \cite{P} is also a special case
of the threshold points for the cumulative weight enumerators.

For any random codes over an alphabet
with $q$ (not necessarily equal to~$2$) elements
we consider a similar setting and obtain similar
sharp threshold points. As immediate consequences,
we show that $\frac{1}{2}{\rm GV}_q(\delta)$ is a very sharp
threshold point for the density of the codes of rate $r$ whose
relative distance is greater than a given $\delta$;
on the other hand, we point out that
the asymptotic distribution of the relative distances
of the random codes with given rate $r$ is exactly located at
$\delta_{\rm GV}(2r)$, as described in \cite{BF}.

The paper is composed of five sections. After the introduction,
in Section~2 the necessary notations are introduced for quotation.
In Section 3 random linear codes are discussed;
while in Section 4 arbitrary random codes are studied.
Finally we draw conclusions in Section 5.

\section{The KL-divergence and the GV-bound}

Throughout the paper, $F$ is an alphabet with cardinality
$|F|=q>1$, $n$ is a positive integer;
and we always take the following parameters:
\begin{equation}\label{parameters}
 \delta\in(0,\,\delta_0)~~{\rm where}~\delta_0=1-q^{-1},
 \quad d=\delta n;\qquad r\in(0,\,1),\quad k=rn;
\end{equation}
where $d=\delta n$ should be understood, in practical computations,
to be the integer nearest to $\delta n$;
and it is the same for $k=rn$.

In this paper, $\exp(x)=e^x$ denotes the natural
exponential function with the natural base $e$,
while $\exp_q(x)=q^x$ denotes the exponential function
with base~$q$;
and $\log_q x$ denotes the logarithm function to the base $q$,
while $\ln x$ denotes the natural logarithm function.
Recall that, for any $0\le p,p'\le 1$,
the $q$-ary {\em KL-divergence} (or {\em information divergence})
of $p$ to $p'$ is defined as
$$\textstyle
 D_q(p\Vert p')=p\log_q\frac{p}{p'}+(1-p)\log_q\frac{1-p}{1-p'}
$$
with the conventions $0\log_q 0=0$ and $0\log_q\frac{0}{0}=0$. It
is a convex non-negative function in $p$ with a unique zero point at
$p=p'$, see \cite[\S1.2]{MM}.

We consider the following partial sum of binomials:
\begin{equation}\label{binom}
 \beta_n(\delta)=\sum\limits_{i=0}^{\delta n}
 \binom{n}{i}\delta_0^i(1-\delta_0)^{n-i}
 =\sum\limits_{i=0}^{d}
 \binom{n}{i}\delta_0^i(1-\delta_0)^{n-i}\,,
\end{equation}
where $\binom{n}{i}$ denotes the binomial coefficient.
We have the following estimation which will
play a key role in the asymptotic analysis of random codes:
\begin{equation}\label{for-binom}\textstyle
\beta_n(\delta)=\exp_q\left(-n D_q(\delta\Vert\delta_0)
-\frac{1}{2}\log_q n+\rho_n(\delta)\right)
\end{equation}
with $\rho_n(\delta)$, $n=1,2,\cdots$, being functions on
$\delta$ such that
$$\textstyle
\lim\limits_{n\to\infty}\exp_q\big(\rho_n(\delta)\big)=\frac
{\sqrt{1-\delta}}{(1-\frac{\delta}{\delta_0})\sqrt{2\pi\delta}}.
$$
We may also write
$$\textstyle\beta_n(\delta)=\exp_q\left(-n D_q(\delta\Vert\delta_0)
 -\frac{1}{2}\log_q n+O(1)\right). \eqno(2.3')
$$

The estimation (\ref{for-binom}) is somewhat known,
for example, \cite[Formula (7)]{P} is just
a special case where $q=2$.
We sketch a proof of (\ref{for-binom}) for completeness.

\smallskip{\bf Proof of (2.3).} ~
Let $T_i=\binom{n}{i}\delta_0^{i}(1-\delta_0)^{n-i}$;
in particular,  $T_d=\binom{n}{\delta n}
\delta_0^{\delta n}(1-\delta_0)^{(1-\delta)n}$.
By Stirling's formula,
$n!=(2\pi n)^{\frac{1}{2}}n^n e^{-n+\gamma(n)}$ with
$\frac{1}{12n+1}<\gamma(n)<\frac{1}{12n}$, so we have an estimation:
\begin{equation}\label{T_d} \textstyle
 T_d=\big(2\pi\delta(1-\delta)n\big)^{-\frac{1}{2}}\exp_q
 \big(-nD_q(\delta\Vert\delta_0)-O(\frac{1}{n})\big).
\end{equation}
For any index $i$ with $0<i<d$ we have
$$\textstyle
\frac{T_{i-1}}{T_{i}}
=\frac{i}{n-i+1}\cdot\frac{1-\delta_0}{\delta_0}
<\frac{i+1}{n-i}\cdot\frac{1-\delta_0}{\delta_0}
=\frac{T_{i}}{T_{i+1}}.
$$
For $d=\delta n$ we let
\begin{equation}\label{a} \textstyle
\alpha=\frac{T_{d-1}}{T_{d}}
=\frac{d(1-\delta_0)}{(n-d+1)\delta_0}
=\frac{(\frac{1}{\delta_0}-1)\delta}{1-\delta+\frac{1}{n}};
\end{equation}
further, for $\delta'$ with $0<\delta'<\delta$ and $d'=\delta'n$,
we let
$$\textstyle
\alpha'=\frac{T_{d'-1}}{T_{d'}}
=\frac{(\frac{1}{\delta_0}-1)\delta'}{1-\delta'+\frac{1}{n}}.
$$
Then
$$\textstyle
 T_{d-i}=\frac{T_{d-i}}{T_{d-i+1}}\cdot\frac{T_{d-i+1}}{T_{d-i+2}}
 \cdots\frac{T_{d-1}}{T_{d}}\cdot T_d\le \alpha^i T_d\,,\qquad
 i=0,1,\cdots,d;
$$
and
$$\textstyle
 T_{d-i}=\frac{T_{d-i}}{T_{d-i+1}}\cdot
 \frac{T_{d-i+1}}{T_{d-i+2}}
 \cdots\frac{T_{d-1}}{T_{d}}\cdot T_d\ge \alpha'^i T_d\,,\qquad
 i=0,1,\cdots,d-d'+1;
$$
thus
$$\textstyle
\frac{1-\alpha'^{(\delta-\delta') n}}{1-\alpha'}T_d=
T_d\sum_{i=0}^{d-d'-1}\alpha'^i~\le~
\sum_{i=0}^{d}T_i~\le~ T_d\sum_{i=0}^d\alpha^i
=\frac{1-\alpha^{1+\delta n}}{1-\alpha}T_d\,;
$$
that is,
$$\textstyle
\frac{(1-\alpha)(1-\alpha'^{(\delta-\delta')n})}{1-\alpha'}
\le \frac{\sum_{i=0}^dT_i}{(1-\alpha)^{-1}T_d}
 \le 1-\alpha^{1+\delta n}.
$$
For any $\varepsilon>0$ we can take $\delta'<\delta$ such that
$\frac{\alpha-\alpha'}{1-\alpha'}<\varepsilon$; hence
$\frac{1-\alpha}{1-\alpha'}>1-\varepsilon$;
in this way we can get that
\begin{equation}\label{equiv}
\lim_{n\to\infty}\frac{\sum_{i=0}^dT_i}{(1-\alpha)^{-1}T_d}=1.
\end{equation}
Finally, combining (\ref{T_d}), (\ref{a}) and (\ref{equiv}),
we obtain the estimation (\ref{for-binom}). \qed

\smallskip As an immediate consequence of (2.3$'$),
we get the following lemma.

\medskip{\bf Lemma 2.1.}~ {\it
Let notations be as in (2.1) and (2.2). Then
\begin{equation}\label{q-beta}\textstyle
q^{rn}\beta_n(\delta)=\exp_q\Big(\big(
r-D_q(\delta\Vert\delta_0)\big)n-\frac{1}{2}\log_q n+O(1)\Big).
\end{equation}
In particular,
\begin{equation}\label{limit-q-beta}
\lim\limits_{n\to\infty}q^{rn}\beta_n(\delta)=\begin{cases}
   0, & r \le D_q(\delta\Vert\delta_0);\\[3pt]
  \infty, & r > D_q(\delta\Vert\delta_0); \end{cases}
\end{equation}
the limit is exponentially convergent for every $r\in(0,1)$
except for $r=D_q(\delta\Vert\delta_0)$;
at the exceptional point it converges to $0$ at the
rate of $\frac{1}{\sqrt n}$ going to $0$. }

\medskip
Recall that any ${\bf a}=(a_1,\cdots,a_n)\in F^n$ is said to be
a word of length $n$ over~$F$;
the {\em Hamming distance} between the words ${\bf a}$ and ${\bf a'}$,
denoted by $d_H({\bf a},{\bf a}')$, is defined to be
the number of the indices $i$ where ${\bf a}$ and ${\bf a'}$
have different coordinates $a_i\ne a_i'$.
By $B({\bf a},\delta n)$ we mean the Hamming ball centered at
${\bf a}\in F^n$ with radius $\delta n$.
The volume of the Hamming ball is denoted by
$V_q(n,\delta n)=|B({\bf a},\delta n)|$.
With the notation (\ref{binom}), we have
\begin{equation}\label{ball}\textstyle
 V_q(n,\delta n)=\sum_{i=0}^{\delta n}\binom{n}{i}(q-1)^i=q^n\beta_n(\delta).
\end{equation}
Any non-empty subset $C$ of $F^n$ is called a $q$-ary code $C$ of length $n$
over $F$, any ${\bf c}\in C$ is said to be a codeword.
We can write the cardinality of $C$ as $|C|=q^{R(C)n}$,
where $R(C)=\frac{\log_q(|C|)}{n}$ is called the {\em rate of $C$}.
The minimum distance of $C$, denoted by  $d_H(C)$, is defined as
$d_H(C)=\min\limits_{{\bf c}\ne{\bf c'}\in C} d_H({\bf c},{\bf c'})$;
and the fraction $\frac{d_H(C)}{n}$
is called the {\em relative distance of $C$},
denoted by $\Delta(C)$.

For a given $\delta\in(0,\delta_0)$, a longstanding important question
is the following: what is the highest rate $R(C)$ for
$C\subseteq F^n$ with $\Delta(C)>\delta$? We sketch it as a remark.

\medskip{\bf Remark 2.1.}~
Applying a typical greedy algorithm to select codewords step by step,
we can get a code $C\subseteq F^n$ with $\Delta(C)>\delta$
such that $\bigcup\limits_{{\bf c}\in C}B({\bf c},\delta n)= F^n$,
hence $|C|\cdot V_q(n,\delta n)\ge q^n$;
by Eqns (\ref{ball}) and (2.3$'$) we have that
\begin{equation}\label{GV}\textstyle
q^{R(C)n} \ge \frac{q^n}{V_q(n,\delta n)}=\exp_q\left(
 n D_q(\delta\Vert\delta_0)+\frac{1}{2}\log_q n-O(1)\right)
\end{equation}
where $\frac{q^n}{V_q(n,\delta n)}$ is just the usual
{\em Gilbert-Varshamov bound} (cf. \cite[\S2.8, \S2.9]{HP}),
which is a function in $n$ and $\delta$;
thus
\begin{equation}\label{asymptotic-GV}\textstyle
R(C)\ge D_q(\delta\Vert\delta_0)+\frac{\log_q n}{2n}-O(\frac{1}{n}),
\end{equation}
which implies that

\smallskip\hangindent30pt
$\bullet$~ {\it for large enough $n$, the highest rate of the
codes of $F^n$ whose relative distance is greater than $\delta$
is larger than $D_q(\delta\Vert\delta_0)$.}

\smallskip
A code $C$ is said to be linear if $F$ is a finite field
and $C$ is a subspace of the vector space $F^n$.
For linear codes, applying a similar greedy algorithm
to construct {\em parity check matrices} of linear codes,
we can get the same conclusions as above.

\smallskip
From the asymptotic perspective, the quantity
$D_q(\delta\Vert\delta_0)$ is known as the
{\em asymptotic Gilbert-Varshamov bound}
for $q$-ary codes at the relative distance $\delta$. We denote it by
${\rm GV}_q(\delta)=D_q(\delta\Vert \delta_0)$,
and call it the {\em GV-bound} in short.
In fact, the GV-bound is also related to the so-called
{\em $q$-ary entropy function}
$$H_q(\delta)=\delta\log_q(q-1)-\delta\log_q\delta
  -(1-\delta)\log_q(1-\delta)$$
in the sense that ${\rm GV}_q(\delta)=1-H_q(\delta)$,
see \cite[\S2.10.6]{HP}.

\medskip
As we have pointed out for the KL-divergence, ${\rm GV}_q(\delta)$
is a convex and strictly decreasing function
in $\delta\in(0,\delta_0)$ with
${\rm GV}_q(0)=1$ and ${\rm GV}_q(\delta_0)=0$.

\medskip{\bf Remark 2.2}.~
In the following, for convenience,
we let $r_\delta={\rm GV}_q(\delta)$ for any
$\delta\in(0,\delta_0)$, and $r_0=1$, $r_{\delta_0}=0$;
i.e. $r_\delta$ is the GV-bound at distance $\delta$.
On the other hand,
because of the monotone property of the function ${\rm GV}_q(\delta)$
in $\delta\in(0,\delta_0)$, for any $r\in(0,1)$
there is a unique real number in the interval $(0,\delta_0)$,
which we denote by $\delta_r$,
such that $r={\rm GV}_q(\delta_r)$;
of course, $\delta_0=1-q^{-1}$ as fixed in (\ref{parameters}),
and $\delta_1=0$. The quantity $\delta_r$
is called the {\em GV-distance} at rate~$r$; see
references \cite{BF}, \cite{M}, \cite[\S6.2]{MM}.

\section{Random linear codes}

In this section, we keep the parameters in (\ref{parameters})
and assume that $F$ is a finite field of order $q$,
hence $F^n$ is a vector space over $F$ of dimension $n$.
The Hamming weight $w_H({\bf a})$ for ${\bf a}\in F^n$ is defined by
$w_H({\bf a})=d_H({\bf a},{\bf 0})$,
where ${\bf 0}$ denotes the zero vector.
Any linear subspace $L$ of the vector space $F^n$
is said to be a linear code;
the rate of $L$ is $R(L)=\frac{\dim L}{n}$,
where $\dim L$ denotes the dimension of the linear code $L$;
while the relative distance of $L$ is $\Delta(L)=\frac{w_H(L)}{n}$,
where $w_H(L)=\min\limits_{{\bf 0}\ne{\bf c}\in L} w_H({\bf c})$
is called the {\em minimum weight} of the linear code $L$.

Consider the probability space ${\cal L}_r(F^n)$
whose sample space is the set of all linear codes of
rate $r$ in $F^n$ and whose probability function is equiprobable.
Take $L_r\in{\cal L}_r(F^n)$, i.e. $L_r$ is a random linear code
of length $n$ and rate $r$. Set
$\hat{\cal N}_{L_r}^{\bf 0}(\delta)$ to be the cardinality of
the set of the non-zero codewords whose weight is
at most $\delta n$, i.e.
\begin{equation}\label{NL-delta}
 \hat{\cal N}_{L_r}^{\bf 0}(\delta)=
 \big|(L_r\backslash\{{\bf 0}\})\cap B({\bf 0},\delta n)\big|,
\end{equation}
where $L_r\backslash\{{\bf 0}\}$ denotes the set difference.
In particular, the minimal $\delta$ such that
$\hat{\cal N}_{L_r}^{\bf 0}(\delta)\ge 1$ is just
the relative distance $\Delta(L_r)$ of $L_r$.
We remark that $\hat{\cal N}_{L_r}^{\bf 0}(\delta)
=\sum\limits_{j=1}^{\delta n}{\cal N}_{L_r}^{{\bf 0}}(j)$,
where ${\cal N}_{L_r}^{{\bf 0}}(j)$ stands for the number of
codewords ${\bf c}$ such that $d_H({\bf 0},{\bf c})=j$,
and is called the {\em distance enumerator
with respect to the codeword~${\bf 0}$}; cf.
\cite[\S B]{BF} or \cite[\S6.2]{MM}.
We name $\hat{\cal N}_{L_r}^{\bf 0}(\delta)$
the {\em cumulative weight enumerator}.

For asymptotic analysis,
we are concerned with the fraction
$\frac{\log_q\hat{\cal N}_{L_r}^{\bf 0}(\delta)}{n}$,
which is called the {\em growth rate} of
$\hat{\cal N}_{L_r}^{\bf 0}(\delta)$, and abbreviated by
$R\big(\hat{\cal N}_{L_r}^{\bf 0}(\delta)\big)$; i.e.
\begin{equation}\label{R-NL-delta}
\hat{\cal N}_{L_r}^{\bf 0}(\delta)=\exp_q
\left(n\cdot R\big(\hat{\cal N}_{L_r}^{\bf 0}(\delta)\big)\right).
\end{equation}
Note that $\hat{\cal N}_{L_r}^{\bf 0}(\delta)$ and
its growth rate are random variables over the probability
space ${\cal L}_r(F^n)$. By Eqn (\ref{R-NL-delta}),
$0\le\hat{\cal N}_{L_r}^{\bf 0}(\delta)\le q^{rn}-1$, so
\begin{equation}\label{RNL-range}
R\big(\hat{\cal N}_{L_r}^{\bf 0}(\delta)\big)
 \in\{-\infty\}\cup[0,r).
\end{equation}

Recalling from Remark 2.2 that $r_\delta={\rm GV}_q(\delta)$
denotes the GV-bound at the distance $\delta$,
we state the main result of this section.

\medskip
{\bf Theorem 3.1.}~ {\it
Let notations be as in (\ref{R-NL-delta}). Assume that
$\delta\in(0,\delta_0)$, $r\in(0,1)$
and $t\in\{-\infty\}\cup[0,r)$. Then
$$\lim\limits_{n\to\infty}
\Pr\left(R\big(\hat{\cal N}_{L_r}^{\bf 0}(\delta)\big)\ge t\right)
=\begin{cases}
   1, & {\rm if}~~ t < r-r_\delta;\\[3pt]
   0, & {\rm if}~~ t\ge r-r_\delta;\end{cases}$$
the limit converges exponentially for any
$t\in\{-\infty\}\cup[0,r)$ except for $t=r-r_\delta$;
at the exceptional point it converges to $0$
at the rate of $\frac{1}{\sqrt n}$ going to $0$.
}

\medskip{\bf Proof.}~
Let ${\cal L}(F^k,F^n)$ be the probability space
whose sample space is the set of all injective linear
maps ${f: F^k\to F^n}$ and
every sample is equipped with equiprobability.
For $f\in{\cal L}(F^k,F^n)$ the image
\begin{equation}\label{Lr}
 L_f=f(F^k)=\{f({\bf b})\mid {\bf b}\in F^k\}\subseteq F^n
\end{equation}
is a linear code of rate $r$ in $F^n$, and any $L_r\in{\cal L}_r(F^n)$
can be obtained in this way; further,
$L_f=L_{f'}$ for $f,f'\in{\cal L}(F^k,F^n)$ if and only if
$f'=fg$ for a $g\in{\rm GL}(F^k)$ where ${\rm GL}(F^k)$ denotes
the set of the linear automorphisms of $F^k$. In a word,
the random linear code $L_f$ and $L_r$ have the same random property
though they are defined in different probability spaces; in particular
$$
\Pr\left(R\big(\hat{\cal N}_{L_r}^{\bf 0}(\delta)\big)\ge t\right)
=\Pr\left(R\big(\hat{\cal N}_{L_f}^{\bf 0}(\delta)\big)\ge t\right).
$$
Thus, instead of $L_r$, we prove the theorem for
the random linear code $L_f$ for $f\in{\cal L}(F^k,F^n)$.
%over the probability space

First we wrtite the random variable
$\hat{\cal N}_{L_f}^{\bf 0}(\delta)$
as a sum of Bernoulli random variables.
For each ${\bf b}\in F^k\backslash\{{\bf 0}\}$,
we have the following indicator random variable
\begin{equation}\label{X0b}
 X_{\bf0, b}=\begin{cases}1, &
  0<w_H\big(f({\bf b})\big)\le\delta n;\\
   0, & {\rm otherwise}. \end{cases}
\end{equation}
Since $f({\bf b})$ runs uniformly over the vectors of
$F^n\backslash\{{\bf 0}\}$
when $f$ runs uniformly over ${\cal L}(F^k,F^n)$,
the expectation of $X_{\bf 0,b}$ is easy to calculate
(cf. Eqn (\ref{ball})):
\begin{equation}\label{EX0b}\textstyle
 {\rm E}\big(X_{\bf 0, b}\big)=\Pr\big(X_{\bf 0, b}=1\big)
 =\frac{V_q(n,\delta n)-1}{q^n-1}=\beta_n(\delta)-O(q^{-n}).
\end{equation}
Further, we construct a non-negative integral random variable
\begin{equation}\label{X0}\textstyle
 X_{\bf 0}=
 \sum\limits_{{\bf b}\in F^k\backslash\{{\bf 0}\}}X_{\bf 0,b}\,.
\end{equation}
It is obvious that $X_{\bf 0}=
 \hat{\cal N}_{L_f}^{\bf 0}(\delta)=\exp_q\left(
  n\cdot R\big(\hat{\cal N}_{L_f}^{\bf 0}(\delta)\big)\right)$,
and $R\big(\hat{\cal N}_{L_f}^{\bf 0}(\delta)\big)\ge t$
is equivalent to $X_{\bf 0}\ge q^{tn}$.

As the second step of the proof, we compute the expectation
of the random variable $X_{\bf 0}$.
By the linearity of expectations and Eqn (\ref{EX0b}) we have
\begin{eqnarray*}
\textstyle{\rm E}\big(X_{\bf 0}\big)&=&
\sum\limits_{{\bf b}\in F^k\backslash\{{\bf 0}\}}
  {\rm E}\big(X_{\bf 0,b}\big)
   =(q^{rn}-1)\big(\beta_n(\delta)-O(q^{-n})\big)\\
  &=&q^{rn}\beta_n(\delta)(1-q^{-rn})-q^{rn}O(q^{-n})(1-q^{-rn});
\end{eqnarray*}
thus, from Eqn (\ref{q-beta}), we obtain the following conclusion
which we write as a lemma for later reference.

\medskip{\bf Lemma 3.1.} ~ {\it
$\textstyle{\rm E}\big(X_{\bf 0}\big)=\exp_q
\big((r-r_\delta)n-\frac{1}{2}\log_qn+O(1)\big)-O(q^{(r-1)n})$;~
in particular,~
$\lim\limits_{n\to\infty}
{\rm E}\big(X_{\bf 0}\big)=\begin{cases}0, & r\le r_\delta;\\
\infty, & r>r_\delta.\end{cases}$
}

\medskip The third step of the proof of Theorem 3.1
is proving the following lemma.

\medskip{\bf Lemma 3.2}.~ {\it If $r>r_\delta$, then
$\lim\limits_{n\to\infty}
\frac{{\rm E}(X_{\bf 0})^2}{{\rm E}(X_{\bf 0}^2)}=1$.}

\medskip{\bf Proof}.~ We fix a ${\bf b_1}\in F^k\backslash\{\bf 0\}$.
For any ${\bf b}\in F^k\backslash\{\bf 0\}$
we have seen from Eqn (\ref{EX0b}) that
${\rm E}(X_{\bf 0,b})={\rm E}(X_{\bf 0, b_1})$; and
it is also obvious that
$${\rm E}(X_{\bf 0}|X_{\bf 0,b}=1)=
 {\rm E}(X_{\bf 0}|X_{\bf 0, b_1}=1),
$$
where ${\rm E}(X_{\bf 0}|X_{\bf 0,b}=1)$ denotes the
conditional expectation of $X_{\bf 0}$ when the event
``$X_{\bf 0,b}=1$'' occurs. By the lemma of \cite[Appendix A]{FS}
we have (or, one can check it directly):
\begin{equation}\label{2-moment}\textstyle
\frac{{\rm E}(X_{\bf 0})^2}{{\rm E}\big(X_{\bf 0}^2\big)}
=\sum\limits_{{\bf b}\in F^k\backslash\{\bf 0\}}
 \frac{{\rm E}(X_{\bf 0,b})}{{\rm E}(X_{\bf 0}| X_{\bf 0, b}=1)}
=\frac{(q^{rn}-1)\cdot {\rm E}(X_{\bf 0, b_1})}
 {{\rm E}(X_{\bf 0}| X_{\bf 0, b_1}=1)}
=\frac{{\rm E}(X_{\bf 0})}{{\rm E}(X_{\bf 0}| X_{\bf 0, b_1}=1)}.
\end{equation}
By the linearity of expectations, we have
$$\textstyle
 {\rm E}(X_{\bf 0}| X_{\bf 0, b_1}=1)
 =\sum\limits_{{\bf b}\in F^k\backslash\{\bf 0\}}
 {\rm E}(X_{\bf 0, b}|X_{\bf 0, b_1}=1).
$$
For ${\bf b}\in F^k\backslash\{{\bf 0}\}$,
there are two cases.

{\it Case 1}:~ ${\bf b}$ and ${\bf b_1}$ are linearly independent.
Since the event ``$X_{\bf 0,b_1}=1$'' occurs,
we consider the injective linear maps $f:F^k\to F^n$ which
satisfy that $f({\bf b_1})\in B({\bf 0},\delta n)$.
When such $f$ is randomly selected, the image
$f({\bf b})$ runs over the vectors in
$F^n\backslash Ff({\bf b_1})$ uniformly at random,
where $Ff({\bf b_1})$ is the $1$-dimensional subspace of $F^n$
generated by $f({\bf b_1})$.
Note that $Ff({\bf b_1})\subseteq B({\bf 0},\delta n)$
since any non-zero vector of $Ff({\bf b_1})$ has
weight $w_H\big(f({\bf b_1})\big)$. Therefore, we get that
\begin{eqnarray*}
{\rm E}(X_{\bf 0, b}|X_{\bf 0, b_1}=1)&=&
 \Pr(X_{\bf 0, b}=1|X_{\bf 0, b_1}=1)
  =\textstyle\frac{V_q(n,\delta n)-q}{q^n-q}\\
  &<&\textstyle \frac{V_q(n,\delta n)-1}{q^n-1}
   ={\rm E}(X_{\bf 0,b_1})
\end{eqnarray*}
where the last equality follows from Eqn (\ref{EX0b}).

{\it Case 2}:~  ${\bf b}$ and ${\bf b_1}$ are linearly dependent.
Then $f({\bf b})$ and $f({\bf b_1})$ are linearly dependent,
hence $w_H\big(f({\bf b})\big)=w_H\big(f({\bf b_1})\big)$; so
${\rm E}(X_{\bf 0, b}|X_{\bf 0, b_1}=1)=1$.

Since the number of the non-zero ${\bf b}$ which are
linearly dependent on ${\bf b_1}$ is equal to $q-1$,
from Eqn (\ref{2-moment}) we get that
$$\textstyle
\frac{{\rm E}(X_{\bf 0}^2)}{{\rm E}(X_{\bf 0})^2}=
\frac{\sum\limits_{{\bf b}\in F^k\backslash\{\bf 0\}}
 {\rm E}(X_{\bf 0, b}|X_{\bf 0, b_1}=1)}
 {{\rm E}(X_{\bf 0})} <
 \frac{(q-1)+(q^{k}-q){\rm E}(X_{\bf 0,b_1})}{{\rm E}(X_{\bf 0})}.
$$
Note that
$\frac{{\rm E}(X_{\bf 0}^2)}{{\rm E}(X_{\bf 0})^2}\ge 1$
(by Cauchy's inequality), and that
$(q^k-q){\rm E}(X_{\bf 0,b_1})<(q^k-1){\rm E}(X_{\bf 0,b_1})
={\rm E}(X_{\bf 0})$ (by (\ref{X0})).
We get that
$$\textstyle
1\le \frac{{\rm E}(X_{\bf 0}^2)}{{\rm E}(X_{\bf 0})^2}<
\frac{q-1}{{\rm E}(X_{\bf 0})}+1.
$$
Since $r>r_\delta$, by Lemma 3.1 we have that
$\lim\limits_{n\to\infty}{\rm E}(X_{\bf 0})=\infty$;
thus we get the desired conclusion $\lim\limits_{n\to\infty}
\frac{{\rm E}(X_{\bf 0})^2}{{\rm E}(X_{\bf 0}^2)}=1$.
\qed

\medskip Now we can complete the proof of Theorem 3.1.

Recall that $\Pr\left(
R\big(\hat{\cal N}_{L_G}^{\bf 0}(\delta)\big)\ge t\right)=
\Pr\big(X_{\bf 0}\ge q^{tn}\big)$.
If $t \ge r-r_\delta$,
then, by Markov's inequality and Lemma 3.1, we have that
$$\textstyle
\Pr\left(X_{\bf 0}\ge q^{tn}\right)\le\frac{{\rm E}(X_{\bf 0})}{q^{tn}}
\le\exp_q\left((r-r_\delta-t)n-\frac{1}{2}\log_qn+O(1)\right);
$$
since $r-r_\delta-t\le 0$, we get
$\lim\limits_{n\to\infty}\Pr\left(X_{\bf 0}\ge q^{tn}\right)=0$.

In the case when $t<r-r_\delta$, we write
$q^{tn}=\theta\cdot{\rm E}(X_{\bf 0})$ where
$$\textstyle
\frac{1}{\theta}=\frac{{\rm E}(X_{\bf 0})}{q^{tn}}=
\exp_q\left((r-r_\delta-t)n-\frac{1}{2}\log_q n+O(1)\right)
-O\big(\exp_q((r-t-1)n)\big);
$$
since $r-r_\delta-t>0$ and $r-r_\delta-t>r-r_\delta-1$,
we have $\lim\limits_{n\to\infty}\theta=0$.
By the Paley-Zygmund inequality (see \cite{PZ})
and Lemma 3.2, we get that
$$\textstyle
\lim\limits_{n\to\infty}\Pr\left(X_{\bf 0}\ge q^{tn}\right)
=\lim\limits_{n\to\infty}\Pr\big(X_{\bf 0}
  \ge\theta{\rm E}(X_{\bf 0})\big)
\ge\lim\limits_{n\to\infty}
(1-\theta)^2\frac{{\rm E}(X_{\bf 0})^2}
{{\rm E}(X_{\bf 0}^2)}=1.
$$

Finally, in all the above arguments
(including the arguments for Lemma 3.2),
the limits are exponentially convergent
except for $t=r-r_\delta$;
at the exceptional point, the limit in Eqn (\ref{limit-q-beta})
of Lemma 2.1 converges to $0$
at the rate of $\frac{1}{\sqrt n}$ going to $0$.
We are done for the proof of Theorem 3.1.\qed

\medskip
Note that $r-t\in(0,r]\cup\{\infty\}$ for $t\in\{-\infty\}\cup[0,r)$.
Since ${\rm GV}_q(\delta)$ is a strictly decreasing
function for $\delta\in[0,\delta_0]$,
with the notations in Remark 2.2 we have:
$$
t<r-r_\delta \iff r_{\delta}<r-t \iff \delta>\delta_{r-t}\,;
$$
$$
t\ge r-r_\delta \iff r_{\delta}\ge r-t \iff \delta\le\delta_{r-t}\,;
$$
where $\delta_{r-t}\in[0,\delta_0)$ with
the following convention:
\begin{equation}\label{delta_infty}
 \delta_{r-t}=0\quad\mbox{if}\quad r-t=\infty.
\end{equation}
Thus we can rewrite Theorem 3.1 as follows.

\medskip{\bf Theorem 3.1$^*$}.~
{\it Let notations be as in Theorem 3.1.
With the convention in (\ref{delta_infty}) we have that
$$\lim\limits_{n\to\infty}
 \Pr\left(R\big(\hat{\cal N}_{L_r}^{\bf 0}(\delta)\big)\ge t\right)
 =\begin{cases}
   0, & {\rm if}~~ \delta\le\delta_{r-t};\\[3pt]
   1, & {\rm if}~~ \delta>\delta_{r-t};\end{cases}$$
the limit converges exponentially for any
$\delta\in(0,\delta_0)$ except for $\delta=\delta_{r-t}$;
at the exceptional point
it converges to $0$ at the rate of $\frac{1}{\sqrt n}$ going to $0$.
}

\medskip If we take $t=0$ and fix $r$, then the statement
``$R\big(\hat{\cal N}_{L_r}^{\bf 0}(\delta)\big)\ge 0$''
is just saying that
``$\hat{\cal N}_{L_r}^{\bf 0}(\delta)\ge 1$'';
and ``$\hat{\cal N}_{L_r}^{\bf 0}(\delta)\ge 1$''
implies that the relative distance satisfies $\Delta(L_r)\le\delta$.
As a consequence of Theorem 3.1$^*$
we obtain again the main result of~\cite{P} at once as follows.

\medskip{\bf Corollary 3.1.}~
{\it Assume that $r\in(0,1)$ is given and $\delta_r\in(0,\delta_0)$
is the corresponding GV-distance (see Remark 2.2). Then
$$
\lim_{n\to\infty}\Pr\big(\Delta(L_r)\le\delta\big)=
\begin{cases}
0, & \delta\le\delta_r;\\[3pt] 1, & \delta>\delta_r;
\end{cases}$$
the limit converges exponentially for any $\delta\in(0,\delta_0)$
except for $\delta=\delta_r$; at the exceptional point
it converges to $0$ at the rate of $\frac{1}{\sqrt n}$ going to $0$.
}

\medskip{\bf Remark 3.1}.~
For any given $r\in(0,1)$, $\Delta(L_r)$ is a random variable
defined over the probability space ${\cal L}_r(F^n)$;
and $\Pr\big(\Delta(L_r)\le\delta\big)$ is just the
cumulative distribution function of $\Delta(L_r)$.
The above corollary shows that, asymptotically speaking,
the distribution of $\Delta(L_r)$ is
concentrated at the GV-distance $\delta_r$; more precisely,
for arbitrarily small positive real number $\varepsilon$ we have
$$
 \lim_{n\to\infty}\Pr\big(\Delta(L_r)\le\delta_r\big)=0=
 \lim_{n\to\infty}\Pr\big(\Delta(L_r)\ge\delta_r+\varepsilon\big),
$$
which is the main result of \cite{P}.

\medskip We now turn to Theorem 3.1. Recalling that
the growth rate of the cumulative weight enumerator
$R(\hat{\cal N}_{L_r}^{\bf 0}(\delta))$ is a random variable
over ${\cal L}_r(F^n)$, we can view Theorem 3.1 as a
characterization of its distribution,
i.e. for arbitrarily small $\varepsilon>0$ we have
\begin{equation}\label{dist-RNL}
 \lim_{n\to\infty}\Pr\big(R(\hat{\cal N}_{L_r}^{\bf 0}(\delta))
   \le r-r_\delta-\varepsilon\big)=0
 =\lim_{n\to\infty}\Pr\big(R(\hat{\cal N}_{L_r}^{\bf 0}(\delta))
   \ge r-r_\delta\big);
\end{equation}
that is, $R(\hat{\cal N}_{L_r}^{\bf 0}(\delta))$ is
asymptotically distributed at exactly one point $r-r_\delta$.
On the other hand, if we take $t=0$ again but fix $\delta$,
then, since ``$\hat{\cal N}_{L_r}^{\bf 0}(\delta)<1$''
is equivalent to ``$\Delta(L_r)>\delta$'',
we get a consequence from Theorem 3.1:

\medskip
{\bf Corollary 3.2.}~ {\it
Assume that $\delta\in(0,\delta_0)$ is given and $r_\delta\in(0,1)$
is the corresponding GV-bound (see Remark 2.2). Then
$$\lim\limits_{n\to\infty}
 \Pr\big(\Delta(L_r)>\delta\big)=\begin{cases}
   1, & {\rm if}~~ r \le r_\delta;\\[3pt]
   0, & {\rm if}~~ r > r_\delta;\end{cases}$$
the limit converges exponentially for any $r\in(0,1)$
except for $r=r_\delta$; at the exceptional point
it converges to $0$ at the rate of $\frac{1}{\sqrt n}$ going to $0$.
}

\medskip
The above corollary implies that the GV-bound
$r_\delta={\rm GV}_q(\delta)$ is a very sharp
threshold point of the parameter $r$ such that
the density of the linear codes with relative distance
greater than $\delta$ changes
from highly dense to extremely sparse.

\medskip{\bf Remark 3.2.}~
More generally, for any point
$(\delta,r)\in [0,\delta_0]\times[0,1]$ we can ask
whether or not there exist linear codes
$L_r$ of rate $r$ such that $\Delta(L_r)>\delta$.
Our results show an asymptotically probabilistic answer
as follows. The rectangle $[0,\delta_0]\times[0,1]$
can be divided into three areas as shown in Figure 3.1:

\begin{figure}[h]\begin{center}\setlength{\unitlength}{1.3pt}
\begin{picture}(125,125)
\thicklines
\put(10,10){\vector(0,1){110}}\put(10,10){\vector(1,0){110}}
\put(5,5){$0$} \put(3,115){$r$} \put(117,1){$\delta$}
%%%%%%%%
\qbezier(10,110)(18,46)(48,26) \qbezier(48,26)(70,10)(105,10)
\thinlines
\put(26,80){\tiny GV-bound} \put(31,80){\vector(-1,-1){11}}
\put(14,38){\small\bf I:}
\put(13,30){\it\small nearly}\put(13,22){\it\small certain}
%%%%%%%%
\qbezier[70](10,110)(23,70)(58,40) \qbezier[50](58,40)(85,17)(105,10)
\put(18,95){\tiny upper bound} \put(30,95){\vector(-1,-1){9}}
\put(32,48){\small\bf II:}
\put(35,43){\it\tiny nearly}\put(35,40){\it\tiny impossible}
%%%%%%%%
\multiput(10,110)(6,0){16}{\line(1,0){2}} \put(4,107){\scriptsize$1$}
\multiput(105,10)(0,6){17}{\line(0,1){2}} \put(102,3){\scriptsize$\delta_0$}
\put(51,65){\small\bf III: \it\small impossible}
%%%%%%%%
\put(80,45){\small\bf Corollary 3.1}
\put(10,36.5){\line(1,0){95}} \put(100,45){\vector(-1,-1){8}}
\qbezier[20](56,20)(33,20)(10,20)
   \put(55,6){\tiny$\delta$}
   \put(-20,19){\tiny$r_\delta={\rm GV}_{\kern-1ptq}(\delta)$}
\put(63,98){\small\bf Corollary 3.2}
\put(57,10){\line(0,1){100}} \put(70,98){\vector(-1,-1){12}}
\qbezier[15](36,10)(36,18)(36,36)
   \put(34,6){\tiny$\delta_r$}
   \put(6,36){\tiny$r$}
\end{picture}\\ Figure 3.1.\quad
Three areas for the probability of the event ``$\Delta(L_r)>\delta$''
\end{center}\vskip-3mm\end{figure}

{\bf Area I:} surrounded by the $\delta$-axis, the $r$-axis and
the curve $r={\rm GV}_q(\delta)$ (the curve itself included),
where every point $(\delta,r)$ satisfies
$\lim\limits_{n\to\infty}\Pr\big(\Delta(L_r)>\delta\big)=1$;
in other words, ``$\Delta(L_r)>\delta$'' is nearly a certain event.
This fact is well known since Varshamov \cite{V}.

{\bf Area II:} surrounded by the curve $r={\rm GV}_q(\delta)$ and
any upper bound (the dotted curve in Figure 3.1);
for any point $(\delta,r)$ in this area, we have
$\lim\limits_{n\to\infty}\Pr\big(\Delta(L_r)>\delta\big)=0$;
that is, in the probabilistic sense,
``$\Delta(L_r)>\delta$'' is nearly an impossible event.
Note that, for a point $(\delta,r)$ of this area,
linear codes with rate $r$ and with
relative distance~$>\delta$ may exist, e.g. see
the works \cite{TVZ,JV,VW,GZ}, see also Remark 2.1.
However, in the asymptotic sense,
the limits (as $n\to\infty$) of the parameters of
the linear codes in \cite{JV,VW,GZ} are located on the curve
$r={\rm GV}_q(\delta)$, hence in Area I,
while the reference \cite{TVZ} does construct linear codes with
parameters in Area II for the case where $q\ge 49$.

{\bf Area III:} above any upper bound,
where ``$\Delta(L_r)>\delta$'' is an impossible event.
There exist many upper bounds, for example,
five upper bounds are described in \cite[\S2.10]{HP}
(but the exact upper bound is still unknown).
Our result also gives an explanation why
it is so hard to determine the exact boundary of the area
where ``$\Delta(L_r)>\delta$'' is an impossible event:
differently from the GV-bound where a very sharp
phase transition occurs, the process from very small probability
to zero probability is a very flat and calm stream,
hence it seems hard to locate accurately the boundary from small to zero.

\medskip{\bf Remark 3.3.}~
Instead of the injective linear maps from $F^k$ to $F^n$ as in
Eqn~(\ref{Lr}), one can consider the probability space
consisting of all the linear maps $\tilde f: F^k\to F^n$ and
with equiprobability.
This probability space is just the so-called
{\em random linear code ensemble} in Shannon's information theory,
cf. \cite[\S6.6]{MM}. Then,
for the random linear codes $\tilde f(F^k)\subseteq F^n$,
all the results of this section still hold.

\section{Arbitrary random codes}

In this section, the parameters in (\ref{parameters})
are still preserved, but $F$ is just an alphabet.
Let $C_r$ be a random code of rate $r$ in $F^n$, in other words,
we consider the probability space ${\cal C}_r(F^n)$
consisting of all codes of rate $r$ in $F^n$
with equiprobability. The issues investigated in this section are
similar to the last section, but the computations and the
analysis of random dependence are different.

Similarly to Eqn (\ref{NL-delta}),
for ${\bf c}\in C_r$ we define
\begin{equation}\label{NC-delta}\textstyle
 \hat{\cal N}_{C_r}^{\bf c}(\delta)=
 \big|(C_r\backslash\{{\bf c}\})\cap B({\bf c},\delta n)\big|
 =\sum\limits_{j=1}^{\delta n}{\cal N}_{C_r}^{{\bf c}}(j),
\end{equation}
where ${\cal N}_{C_r}^{{\bf c}}(j)$ stands for the number of
codewords ${\bf c}'$ such that $d_H({\bf c},{\bf c}')=j$,
and is called the {\em distance enumerator
with respect to the codeword~${\bf c}$}.
Further, we set
\begin{equation}\label{sum-NC-delta}\textstyle
\hat{\cal N}_{C_r}(\delta)=\frac{1}{2}\sum\limits_{{\bf c}\in C_r}
\hat{\cal N}_{C_r}^{\bf c}(\delta)
=\frac{1}{2}\sum\limits_{{\bf c}\in C_r}\,
\sum\limits_{j=1}^{\delta n}{\cal N}_{C_r}^{{\bf c}}(j),
\end{equation}
which is the number of $2$-subsets $\{{\bf c},{\bf c}'\}$
of $C_r$ (i.e. subsets of $C_r$ with cardinality $2$)
such that $d_H({\bf c},{\bf c}')\le\delta n$.
Thus, we name $\hat{\cal N}_{C_r}(\delta)$ the
{\em cumulative distance enumerator}.
In particular, the minimal $\delta$ such that
$\hat{\cal N}_{C_r}(\delta)\ge 1$ is just
the relative distance $\Delta(C_r)$ of $C_r$.
Similar to Eqn (\ref{R-NL-delta}),
we are concerned with the {\em growth rate}
$R\big(\hat{\cal N}_{C_r}(\delta)\big)
 =\frac{\log_q{\hat{\cal N}_{C_r}(\delta)}}{n}$, i.e.
\begin{equation}\label{R-NC-delta}
\hat{\cal N}_{C_r}(\delta)=\exp_q
\left(n\cdot R\big(\hat{\cal N}_{C_r}(\delta)\big)\right).
\end{equation}
Note that $\hat{\cal N}_{C_r}(\delta)$ and its growth rate
are random variables over the probability space ${\cal C}_r(F^n)$.
From Eqn (\ref{sum-NC-delta}) we have that
$0\le\hat{\cal N}_{C_r}(\delta)\le\frac{q^{rn}(q^{rn}-1)}{2}$;
differently from Eqn (\ref{RNL-range}), the range of
$R\big(\hat{\cal N}_{C_r}(\delta)\big)$ is
\begin{equation}\label{RNC-range}
R\big(\hat{\cal N}_{C_r}(\delta)\big)\in\{-\infty\}\cup[0,2r).
\end{equation}

\medskip
{\bf Theorem 4.1.}~ {\it
Let notations be as in (\ref{R-NC-delta}). Assume that
$\delta\in(0,\delta_0)$, $r\in(0,1)$
and $t\in\{-\infty\}\cup[0,2r)$. Then
$$\lim\limits_{n\to\infty}
 \Pr\left(R\big(\hat{\cal N}_{C_r}(\delta)\big)\ge t\right)
 =\begin{cases}
   1, & {\rm if}~~ t < 2r-r_\delta;\\[3pt]
   0, & {\rm if}~~ t\ge 2r-r_\delta;\end{cases}$$
the limit converges exponentially for any
$t\in\{-\infty\}\cup[0,2r)$ except for $t=2r-r_\delta$;
at the exceptional point
it converges to $0$ at the rate of $\frac{1}{\sqrt n}$ going to $0$.
}

\medskip
Similarly to the proof of Theorem 3.1, instead of $C_r$,
we consider %the random code
\begin{equation}\label{Cg}
C_g=\{g({\bf b})\mid {\bf b}\in F^k\}\subseteq F^n
\end{equation}
for $g\in{\cal C}(F^k,F^n)$, where ${\cal C}(F^k,F^n)$ is the
probability space of all injective maps from $F^k$ to $F^n$
(recall that $k=rn$) with equiprobability, since we have
$$
\Pr\left(R\big(\hat{\cal N}_{C_r}(\delta)\big)\ge t\right)
=\Pr\left(R\big(\hat{\cal N}_{C_g}(\delta)\big)\ge t\right).
$$
We first reconstruct the integral random variable
$\hat{\cal N}_{C_g}(\delta)$ with Bernoulli variables.
For any ${\bf b}\ne{\bf b}'\in F^k$
we have the following indicator random variable
\begin{equation}\label{Ybb'} Y_{\bf b, b'}=\begin{cases}
  1, & d_H\big(g({\bf b}),g({\bf b}')\big)\le\delta n;\\
  0, & {\rm otherwise}. \end{cases}
\end{equation}
Since $g$ is injective, we get the expectation as follows:
\begin{equation}\label{EYbb'}\textstyle
 {\rm E}\big(Y_{\bf b, b'}\big)=\Pr\big(Y_{\bf b,b'}=1\big)
 %=\Pr\big(w_H({\bf b}G)\le\delta n\big)
 =\frac{V_q(n,\delta n)-1}{q^n-1}=\beta_n(\delta)-O(q^{-n}).
\end{equation}

\medskip{\bf Remark 4.1.}~
For ${\bf b}\in F^k$ we can define the integral random variable
\begin{equation}\label{Yb}\textstyle
 Y_{\bf b}=
 \sum\limits_{{\bf b}'\in F^k\backslash\{{\bf b}\}}Y_{\bf b, b'}
\end{equation}
so that $Y_{\bf b}=\hat{\cal N}_{C_g}^{g({\bf b})}(\delta)$.
Similar to what we did in Section 3,
one can check that the conclusions of Lemma 3.1, Lemma 3.2
and Theorem 3.1 are all still true for $Y_{\bf b}$, i.e.

(i).~ ${\rm E}(Y_{\bf b})=\exp_q\left(
(r-r_\delta)n-\frac{1}{2}\log_qn -O(q^{(r-1)n})\right)$;

(ii).~ if $r>r_\delta$, then $\lim\limits_{n\to\infty}
  \frac{{\rm E}(Y_{\bf b})^2}{{\rm E}(Y_{\bf b}^2)}=1$;

(iii).~ $\lim\limits_{n\to\infty}
\Pr\left(R\big(\hat{\cal N}_{C_g}^{g({\bf b})}(\delta)\ge
 q^{tn}\big)\right)
=\begin{cases} 1, & t<r-r_\delta;\\
   0, & t\ge r-r_\delta. \end{cases}$

\medskip
However, in order to characterize $\hat{\cal N}_{C_g}(\delta)$,
we need to introduce further an integral random variable:
\begin{equation}\label{Y}\textstyle
Y=\frac{1}{2}\sum\limits_{{\bf b}\in F^k}Y_{\bf b}
 =\sum\limits_{{\bf b}\ne{\bf b}'\in F^k} Y_{\bf b,b'}\,,
\end{equation}
where $\sum\limits_{{\bf b}\ne{\bf b}'\in F^k}$ means that $\{{\bf b},{\bf b}'\}$
runs over the $2$-subsets of $F^k$.
Then we have that $Y=\hat{\cal N}_{C_g}(\delta)
 =\exp_q\left(n\cdot R\big(\hat{\cal N}_{C_g}(\delta)\big)\right)$,
and the event ``$R\big(\hat{\cal N}_{C_g}(\delta)\big)\ge t$''
is equivalent to the event ``$Y\ge q^{tn}$''.

\medskip{\bf Lemma 4.1}. ~ {\it $\textstyle{\rm E}(Y)=\exp_q
\big((2r-r_\delta)n-\frac{1}{2}\log_qn+O(1)\big)-O(q^{(2r-1)n})$;
in particular, $\lim\limits_{n\to\infty}{\rm E}(Y)=\begin{cases}
0, & 2r\le r_\delta;\\ \infty, & 2r>r_\delta.\end{cases}$}

\medskip{\bf Proof}.~ Fix a $2$-subset $\{\bf b_1,b_2\}$
of $F^k$. Let $K=\frac{1}{2}q^{rn}(q^{rn}-1)$.
Let $\{{\bf b},{\bf b}'\}$ be an arbitrary $2$-subset of $F^k$.
By (\ref{EYbb'}) we have
${\rm E}(Y_{\bf b,b'})={\rm E}(Y_{\bf b_1,b_2})$; further,
by the linearity of expectations and Eqn (\ref{Y}), we have
\begin{equation*}
\textstyle{\rm E}(Y)
=K\cdot{\rm E}(Y_{\bf b_1,b_2})
=K\cdot\big(\beta_n(\delta)-O(q^{-n})\big).
\end{equation*}
As $K\beta_n(\delta)=\frac{1}{2}q^{2rn}\beta_n(\delta)(1-q^{-rn})$,
the first equality of the lemma follows from Lemma~2.1.
Then, it is clear that
$\lim\limits_{n\to\infty}{\rm E}(Y)=0$ if $2r\le r_\delta$;
on the other hand, since $2r-r_\delta>2r-1$, we have that
$\lim\limits_{n\to\infty}{\rm E}(Y)=\infty$ if $2r>r_\delta$.
\qed

\medskip{\bf Lemma 4.2}.~ {\it If $2r>r_\delta$, then
$\lim\limits_{n\to\infty}
\frac{{\rm E}(Y)^2}{{\rm E}(Y^2)}=1$.}

\medskip{\bf Proof}.~ Keep the notations as in the proof of Lemma 4.1.
Besides the equality
${\rm E}(Y_{\bf b,b'})={\rm E}(Y_{\bf b_1,b_2})$,
it is also obvious that
${\rm E}(Y|Y_{\bf b,b'}=1)={\rm E}(Y|Y_{\bf b_1,b_2}=1)$.
Similar to the proof of Lemma 3.2, we have:
\begin{equation}\label{2-moment'}\textstyle
\frac{{\rm E}(Y)^2}{{\rm E}\big(Y^2\big)}
=\sum\limits_{{\bf b}\ne{\bf b}'\in F^k}
 \frac{{\rm E}(Y_{\bf b.b'})}{{\rm E}(Y| Y_{\bf b,b'}=1)}
=\frac{K\cdot {\rm E}(Y_{\bf b_1,b_2})}
 {{\rm E}(Y| Y_{\bf b_1,b_2}=1)}
=\frac{{\rm E}(Y)}{{\rm E}(Y| Y_{\bf b_1,b_2}=1)},
\end{equation}
and
$$\textstyle
 {\rm E}(Y|Y_{\bf b_1,b_2}=1)
 =\sum\limits_{{\bf b}\ne{\bf b}'\in F^k}
 {\rm E}(Y_{{\bf b},{\bf b}'}|Y_{\bf b_1,b_2}=1).
$$
For an arbitrary $2$-subset $\{\bf b,b'\}$ of $F^k$,
there are three cases.

{\it Case 1}:~ $\{\bf b,b'\}=\{\bf b_1,b_2\}$,
then ${\rm E}(Y_{\bf b,b'}|Y_{\bf b_1,b_2}=1)=1$.

{\it Case 2}:~ $\{\bf b,b'\}$ and $\{\bf b_1,b_2\}$
have one vector in common, the number of such $2$-subsets is $2(q^{rn}-2)$; for this case, we have
$$\textstyle
 {\rm E}(Y_{\bf b,b'}|Y_{\bf b_1,b_2}=1)=
 \frac{V_q(n,\delta n)-2}{q^n-2}\le\frac{V_q(n,\delta n)-1}{q^n-1}
 ={\rm E}(Y_{\bf b_1,b_2}).
$$

{\it Case 3}:~ $\{\bf b,b'\}\cap\{\bf b_1,b_2\}
=\emptyset$. Since $g$ is injective, we can bound
the conditional expectation as follows
$$
 {\rm E}(Y_{\bf b,b'}|Y_{\bf b_1,b_2}=1)
 \le\frac{V_q(n,\delta n)-3}{q^n-3}
 \le\frac{V_q(n,\delta n)-1}{q^n-1}
 ={\rm E}(Y_{\bf b_1,b_2}).
$$

By Eqn (\ref{2-moment'}) we have
(recall that $K\cdot{\rm E}(Y_{\bf b_1,b_2})={\rm E}(Y)$):
$$\textstyle
1\le\frac{{\rm E}(Y^2)}{{\rm E}(Y)^2}
=\frac{{\rm E}(Y|Y_{{\bf b}_1,{\bf b}_2}=1)}{{\rm E}(Y)}
\le\frac{1+(K-1){\rm E}(Y_{{\bf b}_1,{\bf b}_2})}{{\rm E}(Y)}
<\frac{1}{{\rm E}(Y)}+1.
$$
Because $2r>r_\delta$, by Lemma 4.1 we have
$\lim\limits_{n\to\infty}{\rm E}(Y)=\infty$.
Thus we get the desired conclusion $\lim\limits_{n\to\infty}
\frac{{\rm E}(Y^2)}{{\rm E}(Y)^2}=1$.
\qed

\medskip Recalling that
``$R\big(\hat{\cal N}_{C_g}(\delta)\big)\ge t$'' is equivalent to
``$Y\ge q^{tn}$'' (see Eqn ({\ref{Y}})),
we prove Theorem 4.1 now.

\medskip{\bf Proof of Theorem 4.1}.~
First assume that $t \ge 2r-r_\delta$.
By Markov's inequality and Lemma 4.1, we have that
$$\textstyle
\Pr\left(Y\ge q^{tn}\right)\le\frac{{\rm E}(Y)}{q^{tn}}
\le\exp_q\left((2r-t-r_\delta)n-\frac{1}{2}\log_qn+O(1)\right);
$$
since $2r-t-r_\delta\le 0$, we get
$\lim\limits_{n\to\infty}\Pr\left(Y\ge q^{tn}\right)=0$.

Next, we assume that $t<2r-r_\delta$. We write
$q^{tn}=\theta\cdot{\rm E}(Y)$ where
$$\textstyle
\frac{1}{\theta}=\frac{{\rm E}(Y)}{q^{tn}}=
\exp_q\left((2r-t-r_\delta)n-\frac{1}{2}\log_q n+O(1)\right)
-O\big(\exp_q((2r-t-1)n)\big);
$$
since $2r-t-r_\delta>2r-t-1$, we have that
$\lim\limits_{n\to\infty}\theta=0$;
by the Paley-Zygmund inequality (see \cite{PZ})
and Lemma 4.2, we get that
$$\textstyle
\lim\limits_{n\to\infty}\Pr\left(Y\ge q^{tn}\right)
=\lim\limits_{n\to\infty}\Pr\big(Y\ge\theta{\rm E}(Y)\big)
\ge\lim\limits_{n\to\infty}
(1-\theta)^2\frac{{\rm E}(Y)^2}{{\rm E}(Y^2)}=1.
$$

Finally, in all the arguments above
(including the arguments for Lemma 4.2),
the limits are exponentially convergent
except for $t=2r-r_\delta$; at the exceptional point,
the limit in (\ref{limit-q-beta}) of Lemma 2.1 converges to $0$
at the rate of $\frac{1}{\sqrt n}$ going to $0$.
We are done for the proof.\qed

\medskip
Note that $2r-t\in(0,2r]\cup\{\infty\}$ for
$t\in\{-\infty\}\cup[0,2r)$.
Since ${\rm GV}_q(\delta)$ is a strictly decreasing
function for $\delta\in[0,\delta_0]$, we have
$$
t<2r-r_\delta \iff r_{\delta}<2r-t \iff \delta>\delta_{2r-t}\,;
$$
$$
t\ge 2r-r_\delta \iff r_{\delta}\ge 2r-t
 \iff \delta\le\delta_{2r-t}\,;
$$
where the GV-distance
$\delta_{2r-t}\in[0,\delta_0)$ with
the following convention:
\begin{equation}\label{2-delta_infty}
 \delta_{2r-t}=0\quad\mbox{if}\quad 1<2r-t.
\end{equation}
Similar to Theorem 3.1$^*$, we can rewrite Theorem 4.1 as follows.

\medskip{\bf Theorem 4.1$^*$}.~
{\it Let notations be as in Theorem 4.1.
With the convention in (\ref{2-delta_infty}) we have that
$$\lim\limits_{n\to\infty}
 \Pr\left(R\big(\hat{\cal N}_{C_r}(\delta)\big)\ge t\right)
 =\begin{cases}
   0, & {\rm if}~~ \delta\le\delta_{2r-t};\\[3pt]
   1, & {\rm if}~~ \delta>\delta_{2r-t};\end{cases}$$
the limit converges exponentially for any
$\delta\in(0,\delta_0)$
except for $\delta=\delta_{2r-t}$; at the exceptional point
it converges to $0$ at the rate of $\frac{1}{\sqrt n}$ going to $0$.
}

\medskip If we take $t=0$ and fix $r$, then the statement
``$R\big(\hat{\cal N}_{C_r}(\delta)\big)\ge 0$'' is none other than
``$\hat{\cal N}_{C_r}(\delta)\ge 1$'';
and ``$\hat{\cal N}_{C_r}(\delta)\ge 1$'' is equivalent to
``$\Delta(C_r)\le\delta$'', we get a straightforward
consequence of Theorem 4.1$^*$ as follows.

\medskip{\bf Corollary 4.1.}~
{\it Assume that $r\in(0,1)$ is given and
$\delta_{2r}\in[0,\delta_0)$ is the GV-distance
at the rate $2r$ with the convention in (\ref{2-delta_infty}). Then
$$
\lim_{n\to\infty}\Pr\big(\Delta(C_r)\le\delta\big)=
\begin{cases}
0, & \delta\le\delta_{2r};\\[3pt] 1, & \delta>\delta_{2r};
\end{cases}$$
the limit converges exponentially for any $\delta\in(0,\delta_0)$
except for $\delta=\delta_{2r}$; at the exceptional point
it converges to $0$ at the rate of $\frac{1}{\sqrt n}$ going to $0$.
}

\medskip{\bf Remark 4.2}.~ Similar to the explanation in
Remark 3.1, $\Delta(C_r)$ is a random variable over the
probability space ${\cal C}_r(F^n)$, and
$\Pr\big(\Delta(C_r)\le\delta\big)$ is its
cumulative distribution function.
The corollary exhibits that the distribution of $\Delta(C_r)$
is asymptotically concentrated at the GV-distance $\delta_{2r}$;
more precisely, for an arbitrarily small positive real number
$\varepsilon$ we have
$$
 \lim_{n\to\infty}\Pr\big(\Delta(C_r)\le\delta_{2r}\big)=0=
 \lim_{n\to\infty}\Pr\big(\Delta(C_r)\ge\delta_{2r}+\varepsilon\big).
$$
The references \cite{BF}, \cite{M} have also described
the distribution, but in terms of the distance enumerator
with respect to any given codeword, and not as precisely as above.

\medskip We turn to Theorem 4.1. Similar to Eqn (\ref{dist-RNL}),
$R(\hat{\cal N}_{C_r}(\delta))$ is a random variable
over ${\cal C}_r(F^n)$, and for an arbitrarily small
$\varepsilon>0$ we have
\begin{equation}\label{dist-RNC}
 \lim_{n\to\infty}\Pr\big(R(\hat{\cal N}_{C_r}(\delta))
   \le 2r-r_\delta-\varepsilon\big)=0
 =\lim_{n\to\infty}\Pr\big(R(\hat{\cal N}_{C_r}(\delta))
   \ge 2r-r_\delta\big);
\end{equation}
that is, $R(\hat{\cal N}_{C_r}(\delta))$ is asymptotically
distributed at exactly one point $2r-r_\delta$.
On the other hand, if we take $t=0$ again but fix $\delta$,
then, since ``$\hat{\cal N}_{C_r}(\delta)=0$
(i.e. $\hat{\cal N}_{C_r}(\delta)<1$)'' is equivalent to
``$\Delta(C_r)>\delta$'',
we get a corollary as follows.

\medskip
{\bf Corollary 4.2.}~ {\it
Assume that $\delta\in(0,\delta_0)$ is given and $r_\delta\in(0,1)$
is the corresponding GV-bound. Then
$$\lim\limits_{n\to\infty}
 \Pr\big(\Delta(C_r)>\delta\big)=\begin{cases}
   1, & {\rm if}~~ r \le \frac{1}{2}r_\delta;\\[3pt]
   0, & {\rm if}~~ r > \frac{1}{2}r_\delta;\end{cases}$$
the limit converges exponentially for any $r\in(0,1)$
except for $r=\frac{1}{2}r_\delta$; at the exceptional point
it converges to $0$ at the rate of $\frac{1}{\sqrt n}$ going to $0$.
}

\medskip{\bf Remark 4.3.}~
The above corollary says that
$\frac{1}{2}r_\delta=\frac{1}{2}{\rm GV}_q(\delta)$
is a very sharp phase transition point
for the density of the codes whose relative distance
is greater than the given number $\delta$.
Moreover, similar to Remark~3.2,
for any point $(\delta,r)$ of the rectangle
$(0,\delta_0)\times(0,1)$ we can consider the asymptotic
behavior of the probability $\Pr\big(\Delta(C_r)>\delta\big)$
for the random code $C_r$. The rectangle can be divided
into three areas by the curve $r=\frac{1}{2}{\rm GV}_q(\delta)$
(not the curve $r={\rm GV}_q(\delta)$!)
and the curve for any upper bound as in Figure 4.1:

{\bf Area I:} surrounded by the $\delta$-axis, the $r$-axis and
the curve $r=\frac{1}{2}{\rm GV}_q(\delta)$ (the curve itself included), where ``$\Delta(C_r)>\delta$'' is nearly
a certain event.

{\bf Area II:} surrounded by the curve
$r=\frac{1}{2}{\rm GV}_q(\delta)$ and
any upper bound (the dotted curve),
where ``$\Delta(C_r)>\delta$'' is a nearly impossible event.

{\bf Area III:} above any upper bound (the dotted curve),
where ``$\Delta(C_r)>\delta$'' is an impossible event.

\medskip
\begin{figure}[h]\begin{center}
\begin{picture}(125,125)
\thicklines\put(10,10){\vector(0,1){110}}\put(10,10){\vector(1,0){110}}
\put(3,115){$r$} \put(117,1){$\delta$}
 \put(102,2){\scriptsize$\delta_0$}
%%%%%%%%
\qbezier(10,60)(18,30)(50,17) \qbezier(50,17)(78,8)(105,10)
\put(82,25){\tiny half GV-bound} \put(81,25){\vector(-2,-1){22}}
\put(19,18){\bf\large I}  \put(24,45){\bf\large II}
\thinlines
\qbezier(10,110)(18,46)(48,26) \qbezier(48,26)(70,10)(105,10)
\put(32,75){\tiny GV-bound} \put(31,75){\vector(-3,-2){11}}
\put(26,33){\scriptsize II$'$}  \put(44,33){\scriptsize II$''$}
%%%%%%%%
\qbezier[40](10,110)(20,70)(55,40) \qbezier[30](55,40)(85,14)(105,10)
\put(30,95){\tiny upper bound} \put(30,95){\vector(-3,-2){11}}
\put(60,58){\bf III}
%%%%%%%%
\multiput(10,110)(6,0){16}{\line(1,0){2}} \put(4,107){\scriptsize$1$}
\multiput(105,10)(0,6){17}{\line(0,1){2}} \put(102,2){\scriptsize$\delta_0$}
\end{picture}\\ Figure 4.1\quad
Three areas for the probability of the event
``$\Delta(C_r)>\delta$''
\end{center}\vskip-3mm\end{figure}

\medskip
{\bf Remark 4.4.}~ For an arbitrary random code $C_r$
of rate $r$ in $F^n$, Area II of Figure 4.1
can be further divided by the curve $r={\rm GV}_q(\delta)$
into two subareas II$'$ and II$''$,
where the subarea II$'$ is surrounded by
the curve $r=\frac{1}{2}{\rm GV}_q(\delta)$
and the curve $r={\rm GV}_q(\delta)$ (the curve itself included),
and the subarea II$''$ is the remaining part of Area II; see Figure~4.1.
The two subareas are distinct from each other
in the following two points of view.

$\bullet$~ For any $(r,\delta)$ in the subarea II$'$,
though ``$\Delta(C_r)>\delta$'' is nearly impossible,
by the greedy algorithm one can construct codes of rate $r$
and relative distance $>\delta$; see Remark 2.1.
On the other hand, the greedy algorithm does not work in the
subarea II$''$.

$\bullet$~ For the variable
$Y_{\bf b}=\hat{\cal N}_{C_r}^{g({\bf b})}(\delta)$
defined in Eqn (\ref{Yb}),
from Remark 4.1(iii) one can, by taking $t=0$, get that
$$\lim\limits_{n\to\infty}
\Pr\big(\hat{\cal N}_{C_r}^{g({\bf b})}(\delta)=0\big)
=\begin{cases} 1, & r\le r_\delta;\\
   0, & r> r_\delta; \end{cases}
$$
where the range $r\le r_\delta={\rm GV}_q(\delta)$
covers Area I and the subarea II$'$.
Comparing it with Corollary 4.2, one can see that
the cumulative effect does change the threshold points.

\medskip{\bf Remark 4.5.}~
Instead of the injective maps from $F^k$ to $F^n$ as in
Eqn~(\ref{Cg}), one can consider the probability space whose
sample space consists of all the maps $\tilde g: F^k\to F^n$
and every sample is equipped with equiprobability;
this probability space is just the so-called
{\em random code ensemble} in Shannon's information theory,
cf. \cite[\S6.1]{MM}. Then,
for the random code $\tilde g(F^k)\subseteq F^n$,
all the results of this section still hold.

\section{Conclusion}

For a random code $C$ and ${\bf c}\in C$,
the number of codewords with distance~$j$ from ${\bf c}$,
denoted by ${\cal N}_C^{\bf c}(j)$
and called the {\em distance enumerator with respect to~${\bf c}$},
has been extensively studied with second moment analysis.
In this paper we have introduced the
{\em cumulative distance enumerators} of random codes,
and investigated their thresholds
by more sensitive second moment methods,
including the Paley-Zygmund inequality.
We have shown that the threshold phenomena
of the relative distance and the GV-bound are
just special extreme cases of the threshold phenomena
of the cumulative distance enumerators.

For a random linear code $L$ of length $n$ and rate $r$
over a finite field with $q$ elements
and a number $\delta$ with $0<\delta<1-q^{-1}$,
we defined $\hat{\cal N}_{L}^{\bf 0}(\delta)
 =\sum_{j=1}^{\delta n}{\cal N}_L^{\bf 0}(j)$ and its growth rate
$R(\hat{\cal N}_{L}^{\bf 0}(\delta))
 =\frac{\log_q(\hat{\cal N}_{L}^{\bf 0}(\delta))}{n}$.
We have proved that the growth rate is asymptotically distributed
at exactly one point $r-r_\delta$,
where $r_\delta$ denotes the GV-bound at~$\delta$,
see (\ref{dist-RNL}).
Two consequences, which are more or less already known,
were directly derived: the relative distance $\Delta(L)$ of $L$
is asymptotically distributed at the GV-distance $\delta_r$;
while the GV-bound $r_\delta$ is a sharp threshold point
for the density of linear codes whose relative
distance is greater than $\delta$.

We consider an arbitrary random code $C$ of length $n$ and
rate $r$ over an alphabet with $q$ elements in a similar
manner. For a number $\delta$ with $0<\delta<1-q^{-1}$,
we defined $\hat{\cal N}_{C}(\delta)=\frac{1}{2}\sum_{{\bf c}\in C}
\sum_{j=1}^{\delta n}{\cal N}_C^{\bf c}(j)$, proved that
the growth rate $R(\hat{\cal N}_{C}(\delta))
 =\frac{\log_q(\hat{\cal N}_{C}(\delta))}{n}$
is asymptotically distributed at exactly one point $2r-r_\delta$,
see (\ref{dist-RNC});
and deduced directly two consequences:
the relative distance $\Delta(C)$ is asymptotically distributed
at the GV-distance $\delta_{2r}$; half the GV-bound
$\frac{1}{2} r_\delta$ is a sharp threshold point
for the density of codes whose relative distance
is greater than $\delta$.

\section*{Acknowledgements}
The research of Y.~Fan is supported by NSFC with grant
number 11271005. The research of S.~Ling and C.~Xing
is supported by the Singapore National Research Foundation Competitive
Research Programme (grant number NRF-CRP2-2007-03).
The research of H. W.~Liu and J.~Shen is supported by
NSFC through grant number 11171370.


\begin{thebibliography}{99}

\bibitem{BF}A. Barg, G.D. Forney,
``Random codes: minimum distances and error exponents'',
{\it IEEE Trans. Inform. Theory}, vol.48, pp.2568-2573, 2002.

\bibitem{FS} Y. Fan, J. Shen, ``On the phase transitions
 of random $k$-constraint satisfaction problems'',
{\it Artificial Intelligence}, vol.175, pp.914-927, 2011.

\bibitem{GZ} P. Gaborit, G. Zemor,
``Asymptotic improvement of the Gilbert-Varshamov bound for linear codes'',
{\it IEEE Trans. Inform. Theory}, vol.54, pp.3865-3872, 2008.

\bibitem{Gal} R.G. Gallager, {\it Information Theory and Reliable
Communication}, Wiley, New York, 1968.

\bibitem{Gil} N.E. Gilbert,
``A comparison of signalling alphabets'',
{\it Bell Sys. Tech. Journal}, vol.31, pp.504-522, 1952.

\bibitem{H} R.W. Hamming, ``Error detecting and error correcting codes'',
{\it Bell Sys. Tech. Journal}, vol.29, pp.147-160, 1950.

\bibitem{HP} W.C. Huffman, V. Pless, {\it Fundamentals of
Error-Correcting Codes}, Cambridge University Press, 2003.

\bibitem{JV} T. Jiang, A. Vardy, ``Asymptotic improvement
of the Gilbert-Varshamov bound on the size of binary codes'',
{\it IEEE Trans. Inform. Theory}, vol.50, pp.1655-1664, 2004.

\bibitem{Mon} A. Montanari, ``The glassy phase of Gallager codes'',
Eur. Phys. J. B, vol.23, pp121-136, 2001.

\bibitem{M} N. Merhaev, ``Relations between random coding
exponents and the statistical physics of random codes'',
{\it IEEE Trans. Inform. Theory}, vol.55, pp.83-92, 2009.

\bibitem{MM} M. M\'ezard, A. Montanari,
{\it Information, Physics and Computation},
Oxford University Press, 2009.

\bibitem{PZ} R.E.A.C. Paley and A. Zygmund,
``A note on analytic functions in the unit circle'',
{\it Proc. Camb. Phil. Soc.} vol.28, pp.266-272, 1932.\\
Or, see http://en.wikipedia.org/wiki/PaleyÐZygmund\_inequality

\bibitem{P}
J.N. Pierce,
``Limit distribution of the minimum distance of random linear codes'',
{\it IEEE Trans. Inform. Theory}, vol.13, pp.595-599, 1967.


\bibitem{S} C.E. Shannon,
``A mathematical theory of communication'',
{\it Bell Sys. Tech. Journal}, vol.27, pp.379-423, 623¨C655, 1948.

\bibitem{TVZ} M.A. Tsfasman, S.G. Vladuts, T. Zink,
``Modular curves, Shimura curves and Goppa codes, better than Varshamov-Gilbert bound'',
{\it Math. Nachrichten}, vol.104, pp.13-28, 1982.

\bibitem{V} R.R. Varshamov,
``Estimate of the number of signals in error-correcting codes (in Russian)'',
{\it Dokl. Acad. Nauk}, vol.117, pp.739-741, 1957.

\bibitem{VW} V. Vu, L. Wu,
``Improving the Gilbert-Varshamov bound for $q$-ary codes'',
{\it IEEE Trans. Inform. Theory}, vol.51, pp.3200-3208, 2005.

\bibitem{W} J. Wolfowitz, ``The coding of messages subject to chance errors'',
{\it Illinois J. Math.}, vol.1, pp.591-606, 1957.

%\bibitem{WPZ} http://en.wikipedia.org/wiki/PaleyÐZygmund\_inequality


\end{thebibliography}
\end{document}